\documentclass[showkeys,showpacs,notitlepage,twocolumn]{revtex4-2}
\usepackage[pdfauthor={Richard J. Mathar},colorlinks=true,citecolor=blue,bookmarksopen=true,pdfpagelayout=OneColumn,pdfkeywords={Astrometry,Ephemerides,Radial Velocity}]{hyperref}

\begin{document}

\title[Observer Velocities from JPL Ephemerides]{A Java Program Generating Barycentric Observer Velocities from JPL Ephemerides.}

\author{Richard J. Mathar}
\homepage{https://www.mpia-hd.mpg.de/~mathar} 
\affiliation{K\"onigstuhl 17, 69117 Heidelberg, Germany} 

\pacs{95.10.Jk, 95.75.Pq, 91.10.Ws}

\date{\today}
\keywords{Ephemeris, Astrometry, Earth Rotation, Radial Velocity}

\begin{abstract}
This works presents a program which computes velocities
of an Earth-bound observatory in the reference frame of the barycenter
of the solar system. It feeds from ephemerides files of the Jet Propulsion
Laboratory to extract the velocity of the geocenter, optionally with corrections
from Earth rotation data of the International Earth Rotation Service,
takes a datum (time and geodetic location) of the observer as parameters,
and processes these data with the program library of the working
group `Standards of Fundamental Astronomy' of the International
Astronomical Union.

The prospective application of the computed velocities is to subtract
their projection onto a  pointing direction from observed velocities
in a step of data reduction of astronomic radial velocities.
\end{abstract}

\maketitle 

\section{Size and Velocity Scales} 

From the perspective of the determination of spectrometric radial velocity
measurements in Astronomy, red or blue shifts measure velocities in the observer's
rest frame
\cite{LindegrenAA401}. The ``known'' component of the relative velocity
between observer and star induced by the motion of the observer
in the solar system is of no essential interest,
so a first step of the data reduction is to remove the contributions of \cite{FerrazLNP683}
\begin{enumerate}
\item

    the Earth's elliptic motion around the sun at approximately $\pm 30$ km/s,
\item
    a monthly rotation of $\pm  12$ m/s of the Earth around the barycenter of the Earth-Moon
    two-body system with a period of $\approx 27.3$ days \cite{RoncoliD32296},
\item
    the daily rotation of the telescope with the Earth crust around the Earth axis, which 
    is up to 400 m/s at the equator and proportional to the cosine of the telescope's latitude,
\item
    small contributions from the Earth polar motion of the order of mm/s,
\item
    tiny contributions from the sea tides of the order of centimeters per day \cite{Bos},
\item
    contributions of 230 km/s from the motion around the Galactic center and associated proper motions 
    \cite{SchonrichMNRAS274,ReidAJ832,ValleeASS362}.
\end{enumerate}
We present a Java program which summarizes the first four of these contributions.

\section{Procedural Steps}\label{sec.proc}
The aim of the program is to start from a convenient presentation of the
datum of the observation, a time and a position on the Earth ellipsoid,
and to superimpose the three rotations around the axes mentioned above
to obtain a velocity in a (quasi) non-rotating frame in the solar-system barycenter.

The steps of the computation in our program
are (for given Earth-based position and time)
\begin{enumerate}
\item
Read the coordinates of the Earth-Moon barycenter from the ephemerides
of the Jet Propulson Laboratory (JPL).
The positions are the original data expressed as expansion coefficients
of Chebyshev Polynomials that have scaled time intervals as their arguments. The velocities
are computed by using the standard formulas for derivatives of the Cheybshev Polynomials.
\item
Read the coordinates of the Moon in the geocentric reference frame from the ephemerides.
\item
Merge the two preceeding positions and velocities into position and velocity
of the Earth center by moving $1/(1+r)$ times the Earth-Moon distance
away from the Earth-Moon barycenter, where $r\approx 81.3$ is the ratio of the
two masses taken from the \texttt{EMRAT} value of the Ephemerides header.
The two velocities are combined non-relativistically because the corrections
would be smaller than one millimeter per second.
\item
Read the heliocenter coordinates from the ephemerides.
\item
Optionally read corrections to the models of the resolutions
of the International Astronomical Union (IAU) concerning
precession and nutation of the Earth axis (direction and lag) from Bulletin B files 
of the International Earth Rotation Service (IERS) \cite{BulletinB,BeldaJG}.
\item
Convert the observer's position to WGS84 coordinates
\cite{NIMA8350,JonesJG76,ZhangJG79}
if it was specified as an observatory code of the 
Minor Planet Center (MPC).
\end{enumerate}
Nutation or libration information stored in the ephemerides files is
not used.

All subsequent steps delegate processing to the SOFA library \cite{IAU_sofa},
which embodies the transformation conventions of 2000 and 2006 \cite{Kaplanarxiv06} 
in an efficient interface:
\begin{enumerate}
\item
Construct the elements of the precession-nutation matrix implementing
the IAU 2006 model of precession and IAU 2000 A model of nutation
given a Terrestrial Time (TT).
\item
Extract the direction of the Celestial Intermediate Pole (CIP) from that matrix.
\item
Move the CIP if the program can read corrections from the Bulletin B files
of the IERS found on the local disks.
\item
Fix the azimuths by computing the locator $s$ of the Celestial Intermediate Origin.
\item
Obtain a value of UT1 by an intermediate transformation of TT
to International Atomic Time (TAI) \cite{HohenkM48}. At that point accounting for
leap seconds depends on having installed recent SOFA libraries which
look them up in a hard-coded table of these events.
If the time is too early to define UT1 because UTC was undefined then, the program assumes
that TAI and UTC are the same value.
\item
Compute the Earth Rotation Angle for that UT1.
\item
Obtain the ``locator'' $s'$ of the Terrestrial Intermediate Origin (TIO)
associated with TT.
\item 
In a call of a single function of the library
\begin{enumerate}
\item
Construct the matrix that transforms between the Celestial Intermediate Reference
System and the Geocentric Celestial Reference System (GCRS) (at J2000).
\item 
Compute the observer's geocentric position and velocity.
\item
Rotate these 6 values into the GCRS with the aforementioned matrix.
\item
Obtain the observer's coordinates
in the barycenter,  in the International Celestial Reference System (ICRS),
by adding the velocities of Earth center and observer.
\end{enumerate}
\item
Convert the three velocity components to meters per second and print the results.
\end{enumerate}

\section{Software} 
Our full source code of
the software is in the ancillary directory \texttt{anc} and
licensed under the LGPL; click on the \texttt{details} button
of the arXiv web page
and \texttt{download the entire source package}.

\subsection{Compilation} 
\subsubsection{SOFA} 

As mentioned above, the program absolutely rests
on Harrison's Java rendering of the IAU library \cite{IAU_jsofa}.
There is a long and a short way to install the library:
\begin{itemize}
\item
To compile the code yourself,
move to \url{https://github.com/Javastro/jsofa}, click on \texttt{download}, obtain \texttt{jsofa-master.zip},
and unbundle the files with \texttt{unzip jsofa-master.zip}.
Move (or link) the directory \texttt{jsofa-master/src/main/java/org}
into the same directory as our source code; we
have a directory structure with directories \textit{topdir}\texttt{/de/mpg} and \textit{topdir}\texttt{/org/jastronomy}
for example, if \textit{topdir} (for example \texttt{anc}) 
is the top directory of the Java classes.
Move into the top directory and create the file \texttt{jsofa.jar} with\\
\verb+cd +\textit{topdir}
\begin{verbatim}
export JAVA_TOOL_OPTIONS=-Dfile.encoding=UTF8
javac -cp . org/jastronomy/jsofa/*.java
jar cf jsofa.jar org
\end{verbatim}
or with our \texttt{Makefile} via
\begin{verbatim}
make jsofa.jar
\end{verbatim}
if you have the standard tools of Unices.
\item
To obtain the compiled code right away, move to
\url{http://javastro.github.io/jsofa/}, click on \texttt{binary} in the \texttt{Download}
menu, and rename the file to \texttt{jsofa.jar}.
\end{itemize}

\subsubsection{Local Source} 
To compile the source proposed here, call\\
\verb+cd +\textit{topdir}
\begin{verbatim}
javac -cp . de/mpg/mpia/*/*.java
jar cfm jderv.jar de/mpg/mpia/jderv/Manifest.txt \
   de/*/*/*/*.class
\end{verbatim}
or
\begin{verbatim}
make jderv.jar
\end{verbatim}
in the top directory.

The backslash at the end of the line indicates that the call is
a single command line in the operating system and broken down into many
lines to fit into this manuscript's printout.

For programmers it may be helpful generate a \texttt{doc}
subdirectory with the documentation of the Application Program Interface
(API) with\\
\verb+cd +\textit{topdir}\\
\verb+ javadoc de/*/*/*/*.java org/*/*/*.java \+
\verb+   -encoding utf8 -Xdoclint:none \+
\verb+   -private -sourcepath . -cp . -d doc+

or with
\begin{verbatim}
make doc
\end{verbatim}

\subsection{Configuration} 
The program uses one mandatory data base and two optional data bases
as described in the next three subsections.
\subsubsection{JPL Ephemerides}
The program needs the ASCII files of at least one version of the
ephemerides of NASA's JPL covering the requested time of observation.
It means at least one of the ASCII files with the wildcard name \texttt{ascp*.4??}
and its header file \texttt{header.4??} (with the same and
occasionally in newer versions a longer suffix with an underscore)
must be copied to a local database directory on the local computer.
We shall refer to this directory as \textit{jpldir} further down.
The files are obtained from
\url{https://ssd.jpl.nasa.gov/ftp/eph/planets/ascii/de4??}
where the two question marks depend on which version is wished.

For standard contemporary use one downloads at least \texttt{ascp1950.432}
and \texttt{header.432\_228} into \textit{jpldir}.
This enables the computation of velocities in the years 1950 to 2050
via the option \texttt{-E 432} in the main program (see Section \ref{sec.use}).

Using DE202 and earlier is possible but erroneous,
because these ephemerides did yet not use the coordinate
system of the International Celestial Reference Frame (ICRF)
which the transformations of the program are based on.

Each call of the programs (see below) reads the ASCII files;
reading the binary files also provided by the JPL or creating intermediate
binary files for speedier execution is not supported.

\subsubsection{International Earth Rotation Service}
The Earth Orientation Data can be obtained by starting from 
\url{https://datacenter.iers.org/products/eop/bulletinb/format_2009/csv/}
and downloading any number of 
the files in the Comma-Separated-Values (CSV) format into the
ephemerides directory \cite{BulletinB}.
Rename the files consistently to \texttt{bulletinb-???.csv}
by replacing the \texttt{.txt} suffix
by the \texttt{.csv} suffix.
Alternatively move to \url{https://datacenter.iers.org/products/eop/long-term/c04_14/iau2000/csv/}
and download any set of the yearly files and
rename these files to \texttt{eopc04\_14\_IAU2000.*.csv}.

If the date requested by our program is covered by more than one 
of the  IERS files, the main program takes care to let 
entries marked as \texttt{final} take precedence 
over entries marked as \texttt{predicted}.

These data contain essentially unpredictable corrections to
the Earth Pole position. The files are optional; the main program
assumes that corrections are zero 
if no values for a requested JD (day of observation) can be extracted from the files.
This is the convenient interface definition
for any client interface, because for most applications the values are too small to be of any 
relevance.
The standard reasons for a failure to extract the values are
\begin{enumerate}
\item
The files are not in the current working directory or the parameter provided
by the \texttt{-C} option  was wrong.
\item
The files are present but not readable.
Switching readability on and off (\texttt{chmod a-r bulletinb-*.csv} in Unices)
is therefore a tool to study the impact of these corrections on the velocities.
\item
The JD is outside the union of all days covered by the files on the local
disk. Obviously this occurs if the date is in the future and the extrapolation by the IERS
does not reach that far.
\item
The files are not of the mandatory format with 23 fields separated by semicolons per line, because
they have not been downloaded correctly.
\end{enumerate}

\subsubsection{MPC Observatory Codes}
Observatory codes of the Minor Planet Center (MPC) are supported to
simplify specifying the location of the observer on the Earth
instead of specifying the three parameters in full detail for each call.
That list should be obtained from
\url{https://www.minorplanetcenter.net/iau/lists/ObsCodes.html};
do \emph{not} grab the HTML version \texttt{ObsCodesF.html}.
It must be moved into the same directory as the ephemerides and IERS files.

\subsection{Use} 
\subsubsection{Syntax Main Program} \label{sec.use} 
After compilation the main program is called as

\verb+java -jar jderv.jar+
\textit{[-E 4??]}
\textit{[-D juliandate]}
\textit{[-C jpldir]}
\textit{[-s samples]}
\textit{[-t timeIntvl]}
\textit{[-P ra dec]}
\textit{[-g long lat alt \textbar +g mpccode]}
\textit{[-v]}

The brackets indicate optional arguments and are not to be keyed in.
The vertical bar indicates that only one of the options must be used.
All options and their arguments must be separated by white space.

The options are
\begin{itemize}
\item \texttt{-E} followed by a 3-digit number indicates  the version of
the JPL ephemeris to be used. 
If the option is missing a default is derived
by searching for the environment variable \texttt{PLEPHEM}. If this is the
name (or full path name) of an existing and readable file on the computer, and if
the file part starts with \texttt{header.}, the default number is taken
from the next three digits. If the lookup fails, a default of 432 is assumed.
\item \texttt{-D} provides the time stamp of the first time. 
If
the argument \textit{juliandate} contains the letter \texttt{T}, it must specify a UTC time
stamp in the ISO 8601 format  without time zone designator,
\textit{YYYY}-\textit{MM}-\textit{DD}\texttt{T}\textit{HH}\texttt{:}\textit{mm}\texttt{:}\textit{ss[.ss]}.
Note that the \texttt{T}, the two dashes and two colons, and the specification of the year by 4 digits
and of the month, day, hours and minutes as 2 digits are mandatory in that format.

Otherwise, if the argument is a simple number,
the argument is interpreted as a Julian Date of the Terrestrial Time (TT)
if larger than 2400000, or a Modified Julian Date of TT if smaller than 2400000. 

If the option \texttt{-D} is not used, the time when the program
is called is assumed.

\item \texttt{-C}
The argument of this option indicates the search directory for the ephemerides
files, the optional bulletins B, and the optional \texttt{OpsCodes.html}.
If the option is used, all of these files must be in the same directory---with the aid of
(symbolic) links if needed. The argument must point to a single directory.
This specification may be a relative path name; the option \texttt{-C ..} for
example tells the program to look into the parent of the current directory
of the caller.

If the option is missing a default for the ephemerides directory is derived
by searching for the environment variable \texttt{PLEPHEM}. If this is the
name (or full path name) of an existing and readable file or directory on the computer,
the default string is taken from the directory portion.
If the option \texttt{-C} is not used and this lookup via \texttt{PLEPHEM} fails, 
the default is the current directory of the caller.

If the option is missing a default for the IERS directory is derived
by searching for the environment variable \texttt{iers\_dir}. If this is the
name (or full path name) of an existing and readable directory on the computer,
the directory name is taken from the environment variable.
If the option is not used and this lookup via \texttt{iers\_dir} fails, 
the default is to take the same directory as the ephemerides.

\item \texttt{-s}
Specifies the number of samples on the time axis. \texttt{-s 1} means
that the calculation is performed only once at the date clamped by
the option \texttt{-D}. The option helps to run the program efficiently,
because most of the computer time is spent converting the Ephemerides to a local
binary representation, and this needs to be done only once if the range of
julian dates is covered by a single JPL file.  The default is 100.
\item \texttt{-t}
Specifies the time between the samples in seconds. The default is 60.
\item \texttt{-P}
Specifies a pointing direction by two angles, a right ascension and a
declination. Both are floating point numbers; the right ascension
in hours (modulo $24$) and the declination in degrees in the range $-90$ to $90$.
They may also be submitted in the colon-separated \textit{HH:MM:SS.ss}
and $\pm$\textit{DD:MM:SS.ss} format.

The effect of using that option is that the three Cartesian components
of the computed velocities are projected into that direction and that
speed is printed in two additional columns in the output.
The barycentric corrections of the star position are not taken
into consideration \cite{HrudkovaWDS06,WrightPASP126}, assuming that
the pointing also refers to J2000 coordinates, so the angle between the velocity
vector and the star and therefore the dot product would not change by rotating both to the date of the observation.
\item \texttt{-g}
The three floating point arguments specify the observer's position in
the WGS84: the longitude in degrees in the range $-180$ to $180$, the geographic latitude in 
degrees in the range $-90$ to $90$, and the altitude relative to the
ellipsoid in meters.
\item \texttt{+g}
This option followed by a 3-letter code of the Minor Planet Center (MPC)
specifies the observer's location on Earth (that is, in the ITRF). This
is a lazy way to provide the same information as \texttt{-g} for
the general positions by using the
\texttt{Obscode.html} file as a lookup table.

If the options \texttt{-g} and \texttt{+g} are both used,
\texttt{+g} takes precedence, so both options are effectively exclusive.

If neither \texttt{-g} nor \texttt{+g} is used,
the positions of La Silla default \cite{AngladaAJSS200}.

There is one special flag: \texttt{+g 500} triggers that 
any computations with respect to the observer's
motion with the Earth crust are skipped. The output reflects solely the velocity
of the Earth center in the Solar System Barycenter read off the ephemerides.

Caution: some positions in the \texttt{ObCodes.html} are known only with 4 digits of precision,
which may lead to obscure altitudes if transformed to the WGS84 system. In case of doubt
run the program with \texttt{-v} to obtain the equivalent coordinates, or use the 
\texttt{-g} switch to provide coordinates to higher precision.

\item \texttt{-v}
This option increases the verbosity level and lets the program add
values of intermediate variables into the output stream. This additional
information starts in lines with the hash (\texttt{\#}).
The option helps in particular to watch whether the IERS support data have
actually been inserted into the evaluation or not; if not they are printed
as zeros.

\end{itemize}
The program prints for each time stamp
\begin{enumerate}
\item
the Julian Date (TT),
\item the three Cartesian components of the observer's velocity in the
  coordinate system centered at the solar system barycenter,
\item the modulus of that barycentric velocity,
\item the projected velocity if the \texttt{-P} option was used,
(the dot product of the velocity and the
unit vector into the pointing direction),
followed by the modified value including aberration (light time effects) 
deduced by the radial and transverse components along the
pointing direction \cite{StumpffAA144},
\item the UTC time at the Julian Date in ISO format.
This field is empty if the time lies earlier than the introduction
of the Coordinated Time, which means, lies before a meaningful definition
of leap seconds.
\end{enumerate}

\subsubsection{Minor Planet Equatorial Coordinates} 
A further program is included which transforms
orbits of the Minor Planet Center to equatorial coordinates
as observed from the Earth center at some spot in time.

The call is

\verb+java -jar jderv.jar de.mpg.mpia.jderv.MpcOrbFile +
\textit{[-E 4??]}
\textit{[-D juliandate]}
\textit{[-C jpldir]}
\textit{[-P ra dec coneas]}
\textit{[-m magni]}
\textit{[-a]}

The brackets indicate optional arguments and are not to be keyed in.
All options and their arguments must be separated by white space.

The options \texttt{-E} and \texttt{-C} have the same meaning as above
and specifiy the JPL ephemerides to be read and a search directory.
The search directory \emph{must} contain the file \texttt{MPCORB.DAT}
of the MPC, decompressed, as available from \url{https://www.minorplanetcenter.net/data}.
This is the database of orbital parameters, names and magnitudes of all
planets recognized by the \texttt{MpcOrbFile} program.

The option \texttt{-D} has the same meaning as above, a time stamp
of the observation in UTC or TT scales, with default 
the time when the program is run.

The option \texttt{-P} followed by three arguments defines
a region of interest in equatorial coordinates.
The first and second argument are right ascension and declination
in the same formats as described above, equivalent
to a telescope pointing axis. The third argument
is a maximum separation from that direction in arcseconds.
Planets which are further away from that pointing direction
than the maximum separation at the time of the observation are not printed.
If the option is not used, there is no such filtering 
on positions, which means planets inhabitating
the entire sky (even below the horizon) are considered.

The option \texttt{-m} specifies a limiting magnitude.
If the planet appears to be fainter than that magnitude
at the time of observation, it is not printed.
If the option is not used, the limit is set to 99th magnitude,
which is essentially the same as none.

The option \texttt{-a} indicates that \emph{all} planets
in the \texttt{MPBORB.DAT} file are to be considered,
which are $\approx$ 1.2 million in July 2022.
If the option is not used, the preliminary assignments
in the second part of the file (after a blank line, approximately
half of the planets) are not
considered.

The algorithm of the program is as follows:
The JPL Ephemerides are used to locate the Earth-Moon barycenter
and the relative position Moon-to-Earth
and to superimpose them to yield the position of the Earth 
in the Solar System Barycenter
as described in Section \ref{sec.proc}. Furthermore the position of the sun
is subtracted and the result is a heliocentric position vector of the
center of the Earth.

The \texttt{MPCORB.DAT} file is scanned entirely but chopping off
of the unconfirmed later planets (unless the \texttt{-a} was used).

The following steps are performed for each planet in the
order of appearance in the \texttt{MPCORB.DAT}:

The anomaly $M$ of the planet is moved forward from its ephemerides
time to the observation time linearly with the mean angular motion $n$.
Kepler's equation is solved to derive the angle $E$ \cite[(8.31)(9.18)]{Urban}. The inclination,
major semiaxis and node-perihelion angles yield its
three Cartesian coordinates in the ecliptic plane \cite[(8.32)(8.35)(9.19),(9.21)]{Urban}.
These coordinates are tilted into the equatorial plane (of the epoch J2000)
to produce heliocentric coordinates of the planet.
The vector from the Earth to the Planet is computed by subtraction
of the Earth center.

The light delay is derived by  dividing the distance Earth-to-Planet by the light
velocity,
and the Planet shifted back by that time along its orbit---solving 
once again the Kepler equation. This is not done
to full self-consistency but only once, in tangential order so to speak.

The position vector from the Earth to that light-delay-adjusted
planet position is expressed in the usual right-ascension and declination
angles; if that direction falls outside the cone defined with
the \texttt{-P} option the planet is discarded.

The apparent magnitude is computed from the $H$ and slope-$G$ parameters
via the product of distances to Sun and Earth and mutual
inclination angle  \cite{DymockJBAA117}\cite[(10.38)(10.42)]{Urban}.
If the magnitude is fainter than the limit set by \texttt{-m}
the planet is discarded. At that step planets with a blank
field for $H$ are also considered too faint and discarded.

The program prints for each planet in the \texttt{MPCORB.DAT}
file admitted by the
filtering with the \texttt{-m}, \texttt{-P} and \texttt{-a}
options, separated by commas: 
\begin{itemize}
\item
the MPC designation, 
\item
the right ascension in degrees,
\item
the declination in degrees, 
\item
the apparent visual magnitude,
\item
the light time distance between Earth and planet in seconds,
\item
the right ascension in the hour-minute-seconds format,
\item
the declination in the degrees-arcminutes-arcseconds format,
\item
the long form of the designation, which contains
the unpacked designation in parenthesis and often
the assigned names of the planet.
\end{itemize}

Note that this calculation never enters a stage where
the Earth attitude is needed at times other than
the J2000 epoch. That reference frame is never left.
The intent is to remain comparable with the modern catalogs' standard;
right ascension and declination are not processed
to ``apparent'' positions  including precession-nutation,
aberration or atmospheric refraction effects.

The use case of the program is to provide a warning
for any minor planet that may appear as an unexpected intruder in observations
which look primarily at sidereal targets.
Complementary to the planet-hunters approach to 
find the position of the planet at a given time,
this program acts like a reverse lookup to sift for known planets
\emph{given} position and time.

It is recommended to compare these positions with 
the positions computed by 
\url{https://www.minorplanetcenter.net} under menu \texttt{Observers}
and \texttt{Ephemerides Service}.
This web page offers in addition to take into account small
parallexes induced by the finite distance of the observer
to the Earth center.

The program described here is convenient because one does not
need to know the planets in advance which may happen to appear
in some field of view of the telescope at some particular time.
Its disadvantage is that 7 major planets need to be handled
by other means. (The Earth is obviously not a candidate; Pluto
is included in \texttt{MPCORB.DAT}.) 

To sort the output according to declination one could for example
eliminate the commas and sort numerically by the 3rd field
in the style of:
\verb+java ... | sed 's/,/ /g' | sort -k 3g+.
The alternative is to import the output in a spread sheet program
like \texttt{oocalc} using the comma as separator and to sort inside
that program.

\subsubsection{Syntax Bulletin B Checker} 
A test program that scans the Bulletin B files downloaded to the
local disk is called as

\verb+java -jar jderv.jar de.mpg.mpia.jderv.BulletinB \+\newline
\textit{[-C jpldir]}

The meaning of the \texttt{-C} option is the same as above: it defines
the search directory for files of the form \texttt{bulletinb-*.csv} if
that directory is not the working directory of the caller. This test program
takes a time stamp 90 days backwards from the time of the call, and prints
for 100 days from then on the associated data found in the files. The output
shows the Modified Julian Date (MJD), the $x$ and $y$ components of the CIP offset,
the $dX$ and $dY$ offsets describing polar motion, and the time shift UT1$-$UTC.
The MJD is printed in units of days. The first four values are converted
to radians, so
they differ from the data of the bulletins which are in milli-arcseconds.
The time lag is in units of seconds.

\subsubsection{Examples} 
\begin{verbatim}
java -jar jderv.jar
\end{verbatim}
prints the velocities for an observer on La Silla
based on the DE432 ephemerides (in the current working directory)
for the next 100 minutes in one minute intervals.

\begin{verbatim}
export PLEPHEM=/usr/local/share/jpl/header.430_229
export iers_dir=$HOME/DATA/IERS
java -jar jderv.jar -s 1
\end{verbatim}
prints the velocities for an observer on La Silla
based on the DE430 ephemerides in the \texttt{/usr/local/..} directory
and on the IERS bulletins in the indicated subdirectory of the home directory
observing now.

\begin{verbatim}
java -jar jderv.jar -s 1800 -t 2
\end{verbatim}
prints the velocities for an observer on La Silla
based on the DE432 ephemerides for the next hour in 2 second intervals.

\verb+java -jar jderv.jar -s 1800 \+\newline
\verb:   -t 2 +g 500:

prints the velocities of the Earth center
based on the DE432 ephemerides for the next hour in 2 second intervals.

\verb+java -jar jderv.jar -E 430 \+
\verb:   -s 30 -t 2 +g 000:

prints the velocities of the Greenwich observer
based on the DE430 ephemerides for the next minute in 2 second intervals.

Another test is to compute the Earth barycentric velocities in the  years between 1945 and 1999
once every 500 days
with

\verb+ java -jar jderv.jar -C+ \textit{jpldir} \verb+\+\newline
\verb: -D 2431500.5 -s 41 -t 43200000 +g 500:

The velocity's moduli
differ by less than 0.3 m/s for all years from Stumpff's Table VII \cite{StumpffAAS41}.
The three Cartesian components, however,
differ individually by typically 200 m/s.
To uncover this discrepancy, we derive a mean obliquity of the ecliptic near the middle
of the period, that is 1972 and $T=-0.28$  centuries off the J2000 reference, from
\cite[Eq. (10)]{VondrakAA534}
\begin{eqnarray}
\epsilon_A&\approx& 84028.206305+0.3624445T-0.00004039T^2 \nonumber \\
&& -110\times 10^{-9}T^3 \texttt{arcsec},
\end{eqnarray}
\begin{equation}
\epsilon_A\approx 0.4073797 \texttt{rad}.
\end{equation}
In addition we advance the ascending node at a pace of \cite[Eq. (10)]{VondrakAA534}
\begin{eqnarray}
p_A &\approx& 8134.017132+5043.0520035T-\times 0.00710733T^2 \nonumber \\
&& - 271\times 10^{-9}T^3 \texttt{arcsec},
\end{eqnarray}
over a period of 50 years, $T=0.5$, by $\Delta p_A\approx 0.01222$ rad.
Performing the rotation (precession) with that angle in that ecliptic plane 
---rotation around $x$ by $\epsilon_A$, rotation around $z$ by $\Delta p_A$
and counter-rotation around $x$ by $-\epsilon_A$---reduces the differences
in the velocities' Cartesian coordinates to values of 1.3 m/s or less.
Within that precision the calculation of the program is compatible with Stumpff's models.

\appendix

\section{JPL Reader}
\subsection{Change Log}
The part of our package in the subdirectory \texttt{de/mpg/mpia/jderead}
is a revised version of 
Hristozov's \cite{JDEread}. 
The major differences with respect to the sourceforge version JDEread 1.4
are
\begin{enumerate}
\item
Much of the configuration of the classes that read individual
versions of the ephemerides has been moved into the virtual base
class. In particular all the 
individual astrometric constants are gathered from the ASCII header files;
this avoids typographic errors 
converting these into numbers in the Java code of the derived classes
that occasionally appear in JDEread 1.4.
Also the storage needs are calculated from the
number of Chebychev coefficients read from the header files. 
\item
In consequence the \texttt{header.4??} file associated with
the desired ephemeris version \emph{must} now be present while
the client program reads the main part of the ephemeris.
\item
In consequence the derived classes contain much less code.
Adding readers for additional (forthcoming) editions of the 
ephemerides is reduced to gathering the file names of the ephemerides
and their time spans in the derived classes.
\item
Base classes to read hitherto unsupported ephemerides versions have been added on that
basis.
\item
The API has been changed such that the principal positions that are returned
are measured in kilometers and the associated velocities in kilometers per day.
It is left to the client to divide through the astronomical unit to
get the old units.
\item
A default search path for ephemerides is recognized by scanning
an optional environment variable \texttt{PLEPHEM}.
\item
All indices of array data have been converted from the 1-based
FORTRAN scheme to the 0-based C/C++/Java scheme. The 6-vectors
of positions and velocity have been transformed into $2\times 3$ arrays.
\item
Reading the Chebyshev coefficients is much more based on converting
strings to double with the standard Java libraries than on parsing
mantissae and exponents separately on a term-by-term basis.
\end{enumerate}
\subsection{Standalone Uses}
\subsubsection{Unit Test}
The test program that compares positions and velocities with
the JPL reference values is still available. First download \emph{all}
ephemerides for one type in the time range to be checked, the associated header
file \texttt{header.4??*}, and the reference data \texttt{testpo.4??}
into the directory chosen to keep the data base files. Then call the test program with

\verb+java -cp jderv.jar de.mpg.mpia.jderead.MainTest \ + \newline
\verb+  -E+ \textit{4??} \verb+-C+ \textit{jpldir}

where the number following the option \texttt{-E} is the ephemerides
version to be tested, and where the directory name after the option \texttt{-C}
is the directory with the ephemerides files. If \texttt{-E} is missing
the program will assume 432, and if \texttt{-C} is missing the program
will assume the current working directory.

For DE441 also note that the currently distributed data in the JPL directories
start with file \texttt{ascm13000.441} at ephemerides dates $\approx -3.027\times 10^6$
which is insufficient coverage
of the \texttt{testpo.441} which requires back to $\approx -3.099\times 10^6$.
The coverage in the data files is up $7.93\times 10^6$ in file \texttt{ascp16000.441}
which is also insufficient to run the tests which require data up to $8.00\times 10^6$.
So to run the test successfully the lines 9--2400 and all lines from 362402
on in \texttt{testpo.441} need to be deleted to succeed.

\subsubsection{All-Planets One-Time Snapshot}
The program that gathers positions and velocities for all tabulated
planets at a snapshot in time is called as

\verb+java -cp jderv.jar de.mpg.mpia.jderead.Main \ + \newline
\verb+    -E+ \textit{4??} \verb+-C+ \textit{jpldir} 
\verb+ -D+ \textit{jdate}

where the three options specify available ephemerides version,
directory of the ephemerides files, and a Julian Date as a floating point
number. If the option \textit{-D} is not used, a value of 2440400.5 is assumed,
i.e., the reference value in \cite[Tab. 8.2]{Urban}.

The output shows for each planet the three Cartesian coordinates
of the position in astronomical units (AU), and the three Cartesian coordinates of the
velocity in units of AU per day.  The conversion factor for kilometers per AU
is taken from the parameter \texttt{AU} of the \texttt{header.4??} file of the ephemeris;
it is not a constant.

The enumeration is 
1 for Mercury,
2 for Venus,
3 for the Earth-Moon Barycenter,
4 for Mars,
5 for Jupiter,
6 for Saturn,
7 for Uranus,
8 for Neptun,
9 for Pluto,
10 for the Moon (relative to the Earth center),
and 11  for the Sun.
Some ephemerides have additional data that are shown without scaling
to AU.

\section{MPC Position Checker}
There is a debugging program wired into the parser of the MPC positions
list which can be called as

\verb+java -cp . de.mpg.mpia.jderv.MpcObscod \+\newline
\verb+   -jar jderv.jar+ \textit{jpldir/ObsCodes.html}

where the argument is the full path name of the ASCII file with the stations
list.
This translates all positions from geocentric to geodetic coordinates
and prints for each station the observatory code, the longitude (in degrees),
the geodetic latitude (WGS84, in degrees), the altitude relative to the
ellipsoid (in meters) and the description. Large negative (deep sea) altitudes or altitudes
beyond the few thousand meters of Himalaya summits indicate that
the quality of the entry is dubious
and that these stations ought not be used with the \texttt{+g} flag of the main program.

An equivalent interactive transformation
is implemented in \url{https://dc.zah.uni-heidelberg.de/obscode/q/query/form}.

The program also converts WGS84 positions
into geocentric coordinates of the MPC via option \texttt{-r}
(meaning ``reverse''):

\verb+java -cp . de.mpg.mpia.jderv.MpcObscod \+\newline
\verb+   -jar jderv.jar -r+ \textit{long lat alt}

The final three numbers in this call are the longitude in degrees, the geodetic
latitude in degrees, and the altitude above the ellipsoid in meters.
This can be used to patch inaccurate \texttt{ObsCodes.html} files,
or to attach new codes to the \texttt{ObsCodes.html} file for future
use with the \texttt{+g} option.
A MPC line for the Zugspitze at the German-Austrian border 
could for example be created with

\verb+java -cp . de.mpg.mpia.jderv.MpcObscod \+\newline
\verb+   -jar jderv.jar -r 10.9848606 47.4208929 2965+

\bibliography{eso,all}

\end{document}